\documentclass[11pt]{article}
\usepackage[latin9]{inputenc}
\usepackage[letterpaper]{geometry}
\usepackage[active]{srcltx}
\usepackage{color}
\usepackage{float}
\usepackage{amsthm}
\usepackage{amsmath}
\usepackage{amssymb}
\usepackage{graphicx}
\usepackage[unicode=true,
 bookmarks=true,bookmarksnumbered=true,bookmarksopen=false,
 breaklinks=true,pdfborder={0 0 0},backref=false,colorlinks=true]
 {hyperref}
\hypersetup{pdftitle={Byzantine-Resilient Privacy-Preserving Location-Based Services},
 urlcolor=blue,citecolor=blue,linkcolor=black}

\makeatletter
\usepackage{enumitem}		

\@ifundefined{date}{}{\date{}}
\usepackage{color}
\usepackage{mathrsfs}
\usepackage{amsthm}
\usepackage{breakurl}

\setlist[description]{leftmargin=0pt}

\floatname{algorithm}{Protocol}

\DeclareMathAlphabet{\mathpzc}{OT1}{pzc}{m}{it}

\let\oldenumerate=\enumerate
\def\enumerate{
\oldenumerate
\setlength{\itemsep}{2pt}
\vspace{-4pt}
}

\let\olditemize=\itemize
\def\itemize{
\olditemize
\setlength{\itemsep}{2pt}
\vspace{-3pt}
}

\makeatother

\begin{document}

\title{On Optimal Decision-Making in Ant Colonies}

\author{Mahnush Movahedi\\
 movahedi@cs.unm.edu \and Mahdi Zamani\\
 zamani@cs.unm.edu}

\date{{\small{}Department of Computer Science}\\
{\small{}University of New Mexico, Albuquerque, NM, USA}}
\maketitle
\begin{abstract}
Colonies of ants can collectively choose the best of several nests,
even when many of the active ants who organize the move visit only
one site. Understanding such a behavior can help us design efficient
distributed decision making algorithms. Marshall et al. propose a
model for house-hunting in colonies of ant \emph{Temnothorax albipennis}.
Unfortunately, their model does not achieve optimal decision-making
while laboratory experiments show that, in fact, colonies usually
achieve optimality during the house-hunting process. In this paper,
we argue that the model of Marshall~et~al. can achieve optimality
by including nest size information in their mathematical model. We
use lab results of Pratt~et~al. to re-define the differential equations
of Marshall~et~al. Finally, we sketch our strategy for testing the
optimality of the new model.
\end{abstract}

\section{Introduction}

Understanding \emph{Collective Decision-Making (CDM)} in human and
animal groups can significantly help us design simple and efficient
decentralized algorithms for distributed information systems. Decision
making can be regarded as processing of uncertain information and
producing a final choice among several alternatives. There is a dynamic
tension between the \emph{speed} and the \emph{accuracy} of this process.
The optimality of decision making process is often defined based on
the trade-off between decision accuracy and decision speed. In other
words, an optimal decision is a decision such that no other available
decision options will result in a better trade-off between speed and
accuracy.

The similarities between decision-making mechanisms in human brains
and in colonies of social insects like ants are interesting: both
systems are modeled with mutually interacting populations; in both
systems, a decision is made when the size of one population exceeds
some threshold; and in both systems, this threshold is adjusted to
make a trade off between the decision speed and its accuracy. In the
past two decades, these similarities have attracted the scientific
community to study and model mechanisms of CDM in human and animal
groups~\cite{Visscher07,Passino08,Marshall09}.

To the best of our knowledge, the best known model of biological CDM
is due to Marshall~et~al.~\cite{Marshall09} who compare a model
of decision-making in the primate brain with three models of CDM for
house-hunting of social insect colonies. The first model, proposed
by Pratt~et~al.~\cite{Pratt02}, is related to the emigration of
the rock ant \emph{Temnothorax albipennis}, and the other two models
are related to nest-site selection in the honeybee \emph{Apis mellifera}.
Among the models studied in~\cite{Marshall09}, only direct-switching
model of house-hunting by \emph{A. mellifera} approximates optimal
decision-making in a biologically plausible manner. Unfortunately,
their ant model concludes that colonies of \emph{T. albipennis} cannot
achieve optimal decision-making.

In this paper, we argue that the non-optimal behavior shown by Marshall~et~al.~\cite{Marshall09}
for the house-hunting CDM is not due to the inherent biological deficiencies
but it is due to an unrealistic assumption in~\cite{Marshall09},
where the ants do not consider nest size in their decision. Our goal
is to build a new model of CDM in ant colonies with optimal decision-making
for the house-hunting process. We do this by modifying the model of
\cite{Marshall09} to include nest size information. Similar to~\cite{Marshall09},
we describe our model in terms of stochastic differential equations,
which help us understand the collective behavior in situations with
uncertainty.

As our long-term goal, we envision distributed decision-making algorithms
based on our model for house-hunting in ant colonies. Such algorithms
can be used in several applications including distributed task allocation,
multi-agent systems, supply chain management, and auctions.

\section{Preliminaries}

In this section, we first describe a well-known approach for modeling
CDM called \emph{diffusion model. }Then, we describe the ants house-hunting
CDM algorithm and the mathematical model of Marshall~et~al.~\cite{Marshall09}.
\begin{description}
\item [{Diffusion~Model.}] In this model, decisions are made by a noisy
process that accumulates information over time toward one of the two
alternatives. The model can be thought of as a random walk with normally
distributed step size along a line. Each positive or negative direction
corresponds to increasing evidence for one of the available alternatives.
The random walk is subject to a constant drift, a tendency to move
along the line towards the better alternative, whose strength is the
difference between the expectations of the incoming information on
the available alternatives. The noise in the accumulation of information
is represented as the variance in the random walk.

The diffusion model of decision-making implements \emph{Sequential
Probability Ratio Test (SPRT)} developed by~\cite{wald1945} and
proved it achieves optimal decision-making over two alternatives~\cite{Marshall09},
i.e., by varying the decision threshold it can compromise between
speed and accuracy of decision-making. At the first step, SPRT assume
a pair of hypotheses,$H_{1}:p=p_{0}$ and $H_{2}:p=p_{1}$. The next
step is to gather evidence for the two alternative hypotheses and
calculate the cumulative sum of the log-likelihood ratio, $\log\Lambda_{i}$,
as new data arrive $S_{i}=S_{i-1}+\log\Lambda_{i}$. Finally, the
stopping rule is a simple threshold scheme when the log of the likelihood
ratio exceeds a positive or negative threshold. Through an adjustment
of this threshold, the test can achieve the optimal trade-off between
decision accuracy and speed. The use of log of the likelihood ratio
ensures this test minimizes decision time for any desired decision
error rate.

\item [{Ants~House-Hunting~Algorithm.}] House-hunting of ant \emph{T.
albipennis} has been extensively studied as an interesting CDM behavior~\cite{Pratt02,Conradt05,Pratt05,Frank06}.
Each colony of \emph{T. albipennis} has a single queen and up to about
400 workers as well as brood (eggs, larvae, and pupae) that the colony
has to rear. In the process of house-hunting, scouting ants first
discover new nests and assess them according to some criteria such
as size and darkness. Then, the scouts recruit nest-mates to the new
nest using \emph{tandem-running}, where an informed ant leads a second
ant to her destination to get a second opinion about the nest. When
the number of ants in the new nest reaches a threshold, scouts begin
rapid transport of the rest of the colony by carrying nest-mates and
brood. In each time step of the house-hunting process, each ant is
in one of these three states: \textsf{explore}, \textsf{tandem}, and
\textsf{transport}. The transition between these states happens based
on the ant's evaluation of the quality of the nest sites and the population
of the ants in this sites~\cite{Pratt05}. By conducting laboratory
experiments, Pratt~et~al.~\cite{Pratt05} have observed the following
transition behaviors among \emph{T. albipennis }ants.

\begin{enumerate}
\item If the ant is in the \textsf{explore} state and she has found a new
site, she starts evaluating it. If the site is good, the ant switches
to the state \textsf{tandem}. 
\item If the ant is in the \textsf{tandem} state and the evaluation of the
tandem follower is negative (i.e., the site is not good), the tandem
leader switches to the \textsf{explore} state. If the follower's evaluation
is positive, then with probability $\frac{P^{k}}{P^{k}+T^{k}}$ the
leader switches to the \textsf{transport} state, where $P$ is the
nest population, $T$ is the population at which the probability is
0.5, and $k$ determines the non-linearity of the response, with higher
$k$ yielding a more step-like function~\cite{Pratt05}. 
\item If the ant is in the \textsf{transport} state, she continues evaluating
the site. If she finds a problem with the site, she switches to the
\textsf{explore} state. 
\end{enumerate}

The transition from \textsf{tandem} to \textsf{transport} is interesting
for us since it proves that the number of ants in a site is an important
factor in the ant house-hunting process. We talk about this fact more
in section~\ref{sec:Suggested-Model}.

\item [{House-Hunting~Model~of~Marshall~et~al.}] Marshall~et~al.~\cite{Marshall09}
simplify the house-hunting model of~\cite{Pratt02} by defining the
following factors.

\begin{enumerate}
\item Uncommitted scouting ants $s$, discover nest site $i$ and become
recruiters $y_{i}$ at rate $q_{i}$. This rate is proportional to
the quality of nest size and ease of discovery. Moreover, this rate
is subject to noise $\eta_{q_{i}}$. This is modeled by the term $s\cdot(q_{i}+c\eta_{q_{i}})$
in the equations. 
\item Recruiters $y_{i}$ for site $i$ recruit uncommitted scouts in class
$s$ at a noisy quality-dependent rate $r{}_{i}^{\prime}$ with noise
$\eta_{r{}_{i}^{\prime}}$. This is modeled $y_{i}\cdot(r{}_{i}^{\prime}+\eta_{r{}_{i}^{\prime}})$. 
\item Recruiters for a site $i$ spontaneously switch to recruiting for
the other site $j$ at rate $r_{i}$ subject to noise $\eta_{r_{i}}$.
This is modeled by $y_{j}\cdot(r_{j}+\eta_{r_{j}})-y_{i}\cdot(r_{i}+\eta_{r_{i}})$. 
\item Recruiters $y_{i}$ for a site $i$ spontaneously become uncommitted
to any site at rate $k_{i}$ subject to noise $\eta_{k_{i}}$. This
is modeled by $-y_{i}\cdot(k_{i}+\eta{k_{i}})$. 
\end{enumerate}

For two possible nest sites, the ants CDM process can be described
by the following differential equations 
\[
\left\{ \begin{array}{l}
\dot{y_{1}}=s\cdot(q_{1}+c\eta_{q_{1}})+y_{1}\cdot(r'_{1}+\eta_{r'_{1}})+y_{2}\cdot(r_{2}+\eta_{r_{2}})-y_{1}\cdot(r_{1}+\eta_{r_{1}})-y_{1}\cdot(k_{1}+\eta{k_{1}})\\
\dot{y_{2}}=s\cdot(q_{2}+c\eta_{q_{2}})+y_{2}\cdot(r'_{2}+\eta_{r'_{2}})+y_{1}\cdot(r_{1}+\eta_{r_{1}})-y_{2}\cdot(r_{2}+\eta_{r_{2}})-y_{2}\cdot(k_{2}+\eta{k_{2}})
\end{array}\right.,
\]
where $s=n-y_{1}-y_{2}$. 

\end{description}

\section{Suggested Model \label{sec:Suggested-Model} }

The house-hunting model of Marshall~et~al.~\cite{Marshall09} assumes
that \emph{T. albipennis} ants have no information about their colony
size and the number of ants living in their own nest. In other words,
the ants are unaware of the number of ants who are committed to the
same nest or still uncommitted to any nest and are in the process
of searching or waiting. In contrast, as shown by Pratt~et~al~\cite{Pratt05},
the ants have a sense of these values and can decide based on them.
The new model is described as follows.
\begin{enumerate}
\item \textbf{Dependency to the number of uncommitted scouts.} In~\cite{Marshall09},
recruiters $y_{i}$ switch to the uncommitted state independent of
the number of uncommitted scouts. Also, recruiting uncommitted scouts
in class $s$ is performed independent of the number of uncommitted
scouts. This means that by having more committed scouts to site $i$,
the probability of switching from uncommitted to committed is fixed.
However, we believe this probability decreases until all scouts commit
to the same nest.
\item \textbf{Dependency to the number of scouts in the their current site.}
In~\cite{Marshall09}, recruiters of site $i$ start recruiting for
site $j$ independent of the size of $j$. In contrast, we believe
ants have some idea about their colony size and can observe how many
ants have gathered in one site. In our model, ants use this size information
for committing to nests.
\end{enumerate}
Pratt~et~al.~\cite{Pratt05} argue that the probability of switching
to the \textsf{transport} state is a step function of the normalized
nest population. This probability is almost zero if the number of
ants in the nest is smaller than a threshold, and it is close to one
if this number is larger than a threshold, i.e., 
\[
\Pr(\mathsf{transport})=\left\{ \begin{array}{ll}
0, & \quad\textrm{if }y_{i}\leq T\\
1, & \quad\textrm{if }y_{i}>T
\end{array}\right.,
\]
where $y_{i}$ is the number of ants in site $i$, and $T$ is a fixed
threshold parameter. Since transportation is approximately three times
faster than tandem-running, this step function affects the number
of uncommitted scouts that recruiters $y_{i}$ of site $i$ recruit.
Thus, we can replace the term $y_{i}\cdot(r'_{i}+\eta_{r'_{i}})$
in the model of \cite{Marshall09} with 
\[
\textrm{Number of scouts recruited by recruiters \ensuremath{y_{i}}= }\left\{ \begin{array}{ll}
y_{i}\cdot(r'_{i}+\eta_{r'_{i}}), & \quad\textrm{if }y_{i}\leq T\\
3y_{i}\cdot(r'_{i}+\eta_{r'_{i}}), & \quad\textrm{if }y_{i}>T
\end{array}\right..
\]
Since it is not easy to use this step function directly in our differential
equations, we estimate the function using the following polynomial
computed using Matlab's curve-fitting tool. 
\[
f(y_{i})=82.58y_{i}^{5}-205.33y_{i}^{4}+172.32y_{i}^{3}-54.03y_{i}^{2}+5.71y_{i}-0.11
\]

\begin{figure}[H]
\begin{centering}
\includegraphics[width=0.4\linewidth]{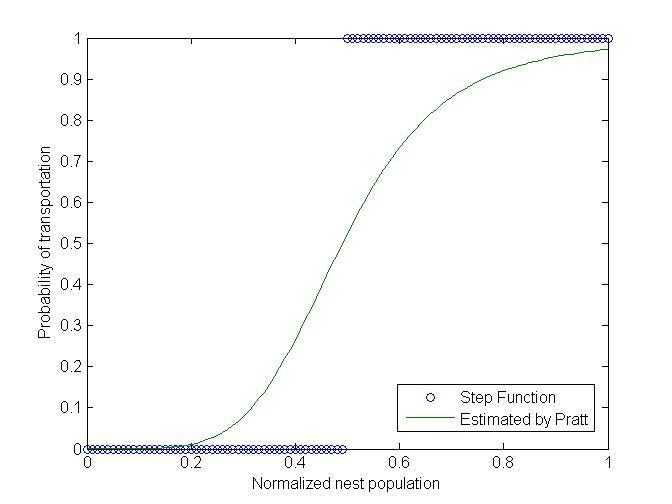} \includegraphics[width=0.4\linewidth]{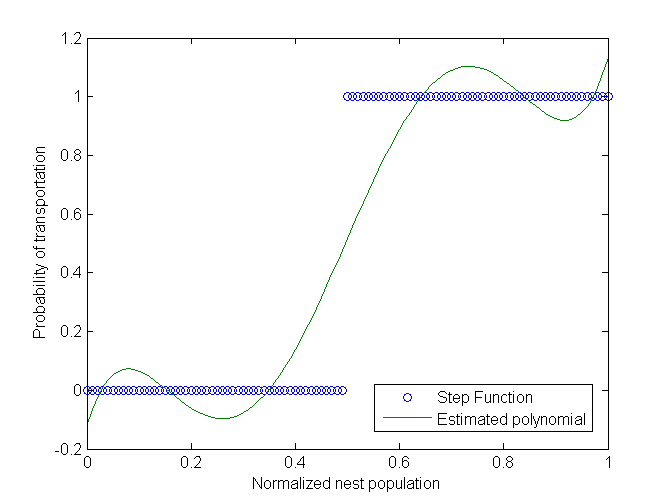}
\protect\caption{Estimation of the step function used by Pratt~et~al.~\cite{Pratt05}
(left) and using a degree-five polynomial (right)}

\par\end{centering}

\label{fig:estimatee} 
\end{figure}

Figure~\ref{fig:estimatee} shows this polynomial and the estimation
used by Pratt~et~al.~\cite{Pratt05} calculated using the probability
$\frac{P^{k}}{P^{k}+T^{k}}$.

We now argue that unlike the model of Marshall~et~al.~\cite{Marshall09},
our model achieves optimal CDM for the house-hunting process among
ants \emph{T. albipennis.} First, we define the model as stochastic
differential equations.
\begin{enumerate}
\item Uncommitted scouts $s$ discover nest site $i$ and become recruiters
$y_{i}$ at rate $q_{i}$. This rate is proportional to the size of
nest and ease of discovery. Moreover, this rate is subject to noise
$\eta_{q_{i}}$. This is modeled as $s(q_{i}+c\eta_{q_{i}})$.
\item Recruiters $y_{i}$ for site $i$ recruit uncommitted scouts in class
$s$ at a noisy quality-dependent rate $r'_{i}$ with noise $\eta_{r'_{i}}$.
This is modeled as $y_{i}\cdot(r'_{i}+\eta_{r'_{i}})\cdot(82.58y_{i}^{5}-205.33y_{i}^{4}+172.32y_{i}^{3}-54.03y_{i}^{2}+5.71y_{i}+0.9)$. 
\item Recruiters for a site $i$ spontaneously switch to recruiting for
the other site $j$ at rate $r_{i}$ subject to noise $\eta_{r_{i}}$.
This is modeled as $y_{j}\cdot(r_{j}+\eta_{r_{j}})-y_{i}\cdot(r_{i}+\eta_{r_{i}})$. 
\item Recruiters $y_{i}$ for a site $i$ spontaneously uncommitted to any
site at rate $k_{i}$ subject to noise $\eta_{k_{i}}$. This is modeled
as $-y_{i}\cdot(k_{i}+\eta{k_{i}})$. 
\end{enumerate}
For two possible nest sites, the ants decision-making process can
be represented by the following equations, 
\begin{equation}
\left\{ \begin{array}{l}
\dot{y_{1}}=s\cdot(q_{1}+c\eta_{q_{1}})+y_{1}\cdot(r'_{1}+\eta_{r'_{1}})\cdot(82.58y_{1}^{5}-205.33y_{1}^{4}+172.32y_{1}^{3}-54.03y_{1}^{2}+5.71y_{1}+0.9)\\
\quad+y_{2}\cdot(r_{2}+\eta_{r_{2}})-y_{1}\cdot(r_{1}+\eta_{r_{1}})-y_{1}\cdot(k_{1}+\eta{k_{1}})\\
\\
\dot{y_{2}}=s\cdot(q_{2}+c\eta_{q_{2}})+y_{2}\cdot(r'_{2}+\eta_{r'_{2}})\cdot(82.58y_{2}^{5}-205.33y_{2}^{4}+172.32y_{2}^{3}-54.03y_{2}^{2}+5.71y_{2}+0.9)\\
\quad+y_{1}\cdot(r_{1}+\eta_{r_{1}})-y_{2}\cdot(r_{2}+\eta_{r_{2}})-y_{2}\cdot(k_{2}+\eta{k_{2}})
\end{array},\right.\label{eq:our-model}
\end{equation}
where $s=n-y_{1}-y_{2}$.

We now briefly describe the steps required for the proof of optimality.
First, we need to transform our model (Equation~\ref{eq:our-model})
into the new system $x_{1}\times x_{2}$, where $x_{1}=\frac{y_{1}-y_{2}}{\sqrt{2}},x_{2}=\frac{y_{1}+y_{2}}{\sqrt{2}}$
and $\dot{x_{1}}=\frac{\dot{y_{1}}-\dot{y_{2}}}{\sqrt{2}},\dot{x_{2}}=\frac{\dot{y_{1}}+\dot{y_{2}}}{\sqrt{2}}$.
Applying these equations to the model gives

\[
\left\{ \begin{array}{l}
\dot{x_{1}}=\frac{n}{\sqrt{2}}(q_{1}+-q_{2}+\sqrt{2}c\eta_{q})+\frac{f_{1}(\frac{\sqrt{}2}{2}(x_{1}+x_{2}))-f_{2}(\frac{\sqrt{}2}{2}(x_{2}-x_{2}))}{\sqrt{2}}\\
\quad+x_{1}\frac{(n-\sqrt{2}x_{2})(r'_{1}+r'_{2})+\sqrt{2}c\eta_{r'}-k_{1}-k_{2}+\sqrt{2}c\eta_{k}-2r_{1}-2r_{2}+2\sqrt{2}c\eta_{r}}{2}\\
\quad+x_{2}\frac{(n-\sqrt{2}x_{2})(r'_{1}-r'_{2})+\sqrt{2}c\eta_{r'}-k_{1}+k_{2}+\sqrt{2}c\eta_{k}-2r_{1}-2r_{2}-2\sqrt{2}c\eta_{r}+2q_{1}-2q_{2}+2\sqrt{2}c\eta_{q}}{2}\\
\\
\dot{x_{2}}=\frac{n}{\sqrt{2}}(q_{1}+q_{2}+\sqrt{2}c\eta_{q})+\frac{f_{1}(\frac{\sqrt{}2}{2}(x_{1}+x_{2}))+f_{2}(\frac{\sqrt{}2}{2}(x_{2}-x_{1}))}{\sqrt{2}}\\
\quad+x_{1}\frac{(n-\sqrt{2}x_{2})(r'_{1}-r'_{2})+\sqrt{2}c\eta_{r'}-k_{1}+k_{2}+\sqrt{2}c\eta_{k}}{2}\\
\quad+x_{2}\frac{(n-\sqrt{2}x_{2})(r'_{1}+r'_{2})+\sqrt{2}c\eta_{r'}-k_{1}-k_{2}+\sqrt{2}c\eta_{k}-2q_{1}-2q_{2}+2\sqrt{2}c\eta_{q}}{2}
\end{array},\right.
\]
where $f_{i}(x)=(r'_{i}+\eta_{r'_{i}})(82.58x^{6}-205.33x^{5}+172.32x^{4}-54.03x^{3}+5.71x^{2})$.
Unfortunately, we cannot simply set the exponent of $x_{1}$ and $x_{2}$
to zero because of the $f_{i}(x)$ terms in the equations. Similar
to~\cite{Marshall09}, it is required to check if this equation can
asymptotically approximate the constant drift diffusion model. To
this end, we need to fix a value for $x_{2}$ independent of $x_{1}$
and then, analyze the behavior of $\dot{x_{1}}$.

\section{Discussion and Conclusion}

We defined new differential equations for the house-hunting process
among ants based on the dependency to the number of scouts in the
nests. We justified our model by referring to the empirical results
of Pratt~et~al.~\cite{Pratt05}. Finally, we sketched our strategy
for testing the optimality of the new model.

One major difference between our model and the model of Marshall~et~al.~\cite{Marshall09}
is that they assume the decision-making process terminates once the
ants start the transportation phase. However, we believe that the
transportation step is an important part of the decision-making process
and can significantly affect the result of the process. Switching
from the \textsf{tandem} state to \textsf{transport} state happens
when the number of ants in the nest is larger than a threshold and
the probability of this transition is a step function which can be
estimated by a polynomial of degree five. At the early stage of the
process, the ants in both nests are in the \textsf{tandem} state.
Once the population of the nest reaches a threshold, the ants may
switch the \textsf{transport} state which can happen in both nests.
This transition from one state to another should not be assumed as
a termination signal because, in fact, the ants may change their decision
and go back to \textsf{tandem} state or even switch to a new nest.

Adding the \textsf{transport} state to the model makes this equations
dependent to the nest size which shows itself in having a polynomial
of degree five in equations. Later on in testing the optimality, this
change shows itself to have terms dependent to both $x_{1}$ and $x_{2}$
and make it impossible to have the random process $\dot{x_{1}}$ independent
to $x_{2}$ and itself which rejects the result of~\cite{Marshall09}.
However, it is the first step in the proof of optimality.

Our main challenge for the future is to complete our proof of optimality,
which we envision to require steps similar to those described in Appendix~D
of~\cite{Marshall09}. We are also interested in checking consistency
of the predictions made by our model to empirical results obtained
from simulations of the house-hunting CDM. Moreover, we are interested
in a probabilistic model of the house-hunting process where the expected
number of ants in each nest is considered. This approach seems useful
for checking if the ants find the best nest with high probability
when the colony has enough time to choose between alternatives.

{\small{}\bibliographystyle{abbrv}
\bibliography{bio}
} 
\end{document}